\documentclass[sigconf]{acmart}

\usepackage{multicol}
\usepackage{subcaption}
\usepackage{amsmath}
\usepackage{enumitem}
\usepackage{multirow}
\usepackage{tabularx}
\usepackage{booktabs}
\usepackage[most]{tcolorbox}
\usepackage{listings}
\setlist[itemize]{noitemsep,topsep=2pt,leftmargin=*}
\setlist[enumerate]{noitemsep,topsep=2pt,leftmargin=*}

\usepackage{fancyvrb}
\fvset{fontsize=\scriptsize,baselinestretch=1}

\newtcolorbox{promptboxcompact}[1]{%
  enhanced,
  colback=white,
  colframe=gray!60,
  boxrule=0.5pt,
  arc=1.5pt,
  left=3pt,right=3pt,top=3pt,bottom=3pt,
  fonttitle=\bfseries\scriptsize,
  title=#1
}
\usepackage{tabularx}
\usepackage{siunitx}
\usepackage{adjustbox}
\usepackage{makecell}
\usepackage{caption}
\usepackage{graphicx}
\usepackage{dblfloatfix} 
\AtBeginDocument{%
  }
\usepackage{siunitx}
\usepackage{microtype}
\usepackage{enumitem}
\usepackage{listings}

\usepackage{caption}
\usepackage{listings}
\usepackage{ragged2e}
\usepackage[most]{tcolorbox} 

\lstdefinestyle{cppstyle}{
  language=C++,
  basicstyle=\ttfamily\footnotesize, 
  columns=fullflexible,
  breaklines=true, breakatwhitespace=true,
  keepspaces=true, showstringspaces=false, tabsize=2
}

\usepackage[most]{tcolorbox} 

\tcbset{
  subtitle style={
    colback=black!5, colframe=black!20,
    boxrule=0.6pt, arc=2mm,
    left=2.5mm, right=2.5mm, top=3pt, bottom=3pt,
    fontupper=\bfseries, coltext=black
  }
}

\newtcolorbox{megabox}[1][]{%
  enhanced, breakable,
  colback=white,
  frame hidden, boxrule=0pt,  
  arc=0pt,                    
  left=2mm, right=2mm, top=2mm, bottom=2mm, boxsep=1.2pt,
  fontupper=\normalsize\RaggedRight,
  fonttitle=\bfseries,
  underlay first={
    \draw[black!45,line width=.6pt] (interior.north west) -- (interior.north east);
    \draw[black!45,line width=.6pt] (interior.north west) -- (frame.south west);
    \draw[black!45,line width=.6pt] (interior.north east) -- (frame.south east);
    \draw[black!45,line width=.6pt] (frame.south west) -- (frame.south east);
  },
  underlay middle={
    \draw[black!45,line width=.6pt] (interior.north west) -- (interior.south west);
    \draw[black!45,line width=.6pt] (interior.north east) -- (interior.south east);
  },
  underlay last={
    \draw[black!45,line width=.6pt] (interior.north west) -- (interior.south west);
    \draw[black!45,line width=.6pt] (interior.north east) -- (interior.south east);
    \draw[black!45,line width=.6pt] (frame.south west) -- (frame.south east);
  },
  #1
}


\renewcommand\footnotetextcopyrightpermission[1]{}

\begin{document}


\title{Beyond Correct Patches: Aligning Code Repair Feedback with Developer Preferences}

\author{Zihan Fang}
\affiliation{
  \institution{Vanderbilt University}
  \city{Nashville}
  \state{TN}
  \country{USA}
}
\email{zihan.fang@vanderbilt.edu}

\author{Yifan Zhang}
\affiliation{
  \institution{Vanderbilt University}
  \city{Nashville}
  \state{TN}
  \country{USA}
}
\email{yifan.zhang.2@vanderbilt.edu}

\author{Yueke Zhang}
\affiliation{
  \institution{Vanderbilt University}
  \city{Nashville}
  \state{TN}
  \country{USA}
}
\email{yueke.zhang@vanderbilt.edu}

\author{Kevin Leach}
\affiliation{
  \institution{Vanderbilt University}
  \city{Nashville}
  \state{TN}
  \country{USA}
}
\email{kevin.leach@vanderbilt.edu}

\author{Yu Huang}
\affiliation{
  \institution{Vanderbilt University}
  \city{Nashville}
  \state{TN}
  \country{USA}
}
\email{yu.huang@vanderbilt.edu}


\begin{abstract}
Large Language Models (LLMs) are increasingly used in software engineering tasks, particularly for code repair.
However, developers often struggle to interpret model outputs, limiting effective human–AI teaming (i.e., humans and AI collaboratively working toward a shared objective). 
Prior work largely focuses on optimizing the generated code, while giving limited attention to the natural-language feedback that supports comprehension and iterative improvement. 
We present \textsc{DPO-f+}, a framework that aligns code-repair feedback with the needs of different developer groups (e.g., novices and proficient developers).
It (1) defines feedback-alignment metrics using seven fixed dimensions with task-specific descriptions; (2) automatically constructs pairwise preference datasets from code-repair tasks; (3) fine-tunes models using Direct Preference Optimization (DPO) augmented with a reward model; and (4) provides an automated protocol for evaluating feedback quality.
Empirically, \textsc{DPO-f+} outperforms both the baseline and standard DPO in terms of feedback accuracy and overall feedback alignment.
On novice programming tasks, \textsc{DPO-f+} improves Pass@1 by 5.71 percentage points (pp) over the baseline and by 3.30 pp over DPO. On the more challenging \textsc{SWE-Bench} benchmark, it improves the issue-resolution rate by 1.67 pp over DPO and by 4.67 pp over the baseline. 
It also significantly improves feedback alignment, as measured by both LLM judges and a human study with 200 developers; in the human study, beginner developers preferred \textsc{DPO-f+} in 71.5\% of comparisons, and overall preference was above chance ($p=0.0057$).
By aligning feedback more closely with developer needs, \textsc{DPO-f+} transforms LLM assist from a one-shot output into a collaborative sense-making workflow. This provides a practical approach to enhance AI-generated code comprehension and to foster more effective human–AI teaming in software engineering.
\end{abstract}



\maketitle

\section{Introduction}
Large Language Models (LLMs) have demonstrated great potential in supporting a wide range of software engineering (SE) tasks, such as code generation, code repair, and automated documentation~\cite{teaming2022state}. 
As LLMs become increasingly integrated throughout the software development lifecycle, human-AI teaming has emerged as a predominant theme in SE. 
In this context, human--AI teaming refers to developers iteratively collaborating with an AI assistant toward a shared objective~\cite{teaming2022state}. We consider teaming effective when developers can readily understand AI-generated outputs and reach successful outcomes with fewer iterations and stronger comprehension.
Task-specific fine-tuning of LLMs for software engineering strengthens human–AI teaming by aligning models with specialized tasks and iterative developer workflows~\cite{jin2024llms}.
Among these applications, code repair has been studied across commonly used settings, such as programming education, open-source maintenance, and AI-assisted pair programming, and has received substantial research attention~\cite{zhang2024pydex,tang2025boosting,zhang2024pair}. 
In these settings, developers commonly submit code to LLMs for correction and iterate on model suggestions. Accordingly, effective human–AI teaming depends on comprehensible feedback rather than code alone, enabling developers to understand the rationale for changes, make targeted edits, and work more efficiently~\cite{lou2025unraveling}.
However, prior work shows that developers often struggle to understand LLM-generated feedback. 
For example, a computer science education study revealed that CS1 students demonstrated a per-task success rate of only 32.5\% in comprehending code and its corresponding explanations generated by LLMs~\cite{zi2025would}. 
Empirical research indicates that for a larger cohort of developers, approximately 20\% of questions on forums are related to understanding content generated by models ~\cite{chen2025empirical}.
Consequently, while many SE studies focus on improving code quality to enhance LLM reliability in repair tasks, the accompanying natural-language feedback is equally important: it helps developers locate errors, understand the rationale behind fixes, and apply repaired code effectively~\cite{pan2024automatically}.
Moreover, since developers’ feedback needs vary by proficiency~\cite{widyasari2025explaining}, profile-specific feedback is necessary to better support human–AI teaming, thereby reducing clarification cycles and improving workflow efficiency.
Together, these highlight the need for alignment frameworks that tailor LLM behavior to diverse developer profiles, prioritizing developer-centered feedback that connects technical code improvements with comprehensible explanations.

Reinforcement Learning from Human Feedback (RLHF) provides a promising approach for adapting code feedback to developer preferences by optimizing language models toward human-preferred outputs~\cite{ouyang2022training}. 
However, Proximal Policy Optimization-based RLHF (PPO) is computationally and annotatively expensive, often requiring online rollouts and large-scale human preference labeling~\cite{ouyang2022training}. These requirements are impractical in resource-constrained settings like computer science education.
By contrast, Direct Preference Optimization (DPO) emerges as a more scalable alternative, enabling direct optimization from offline preference datasets without complex online reinforcement learning, thus streamlining the training process while preserving effective alignment capabilities~\cite{rafailov2023direct}.
However, standard DPO relies on binary pair orderings and ignores the preference margin between candidates~\cite{rafailov2023direct}. 
Moreover, although DPO-inspired methods have shown promise for code generation~\cite{zhang2025focused}, its potential for generating developer-aligned feedback that integrates code edits with tailored explanation remains unexplored in SE, and particularly in code repair settings.

Thus, in this study, we take a first step toward aligning LLM-generated code-repair feedback with developer needs by proposing \textsc{DPO-f+}, a novel and cost-effective framework evaluated across diverse settings. By optimizing profile-specific feedback criteria for different developer groups, our approach also suggests a promising pathway for improving code comprehension with LLMs.
The framework consists of four components: (1) feedback-alignment metrics for assessing alignment, (2) an automated method for constructing pairwise preference datasets to support training, (3) a fine-tuning approach that integrates Direct Preference Optimization (DPO) with a reward, and (4) an automated protocol for evaluating feedback quality.
Empirical results demonstrate that \textsc{DPO-f+} consistently outperforms both baseline and standard DPO. 
It achieves superior feedback accuracy, improving \textit{Pass@1} by 5.71 pp over the baseline and 3.30 pp over standard DPO, while also achieving the highest overall alignment scores as evaluated by both LLM-as-a-judge and 200 developers. 
On more complex tasks (i.e., \textit{SWE-bench Lite}), it improves the resolution rate by 1.67 pp over standard DPO and 4.67 pp over the baseline.
We make the following contributions:
\begin{itemize}
\item A novel framework for alignment of generated feedback in code repair.
\item A reward-augmented DPO that optimizes for feedback under profile-specific criteria without online RL.
\item A feasible approach to improve code comprehension through LLMs in software engineering.
\end{itemize}

Moving beyond accuracy alone, our framework reframes code repair as a collaborative sense-making process that helps developers better understand AI-generated outputs and reduces trial-and-error iterations through feedback tailored to different developer groups. By optimizing LLM-generated feedback, it offers a practical path to improving code comprehension and strengthening human–AI teaming in software engineering.
\begin{figure*}[!htbp]
\caption{Example feedback on the same buggy \texttt{MinStack} implementation from the baseline, standard DPO, and \textsc{DPO-f+}. Colored highlights link each diagnosed issue to its corresponding fix in the suggested patch. The example is randomly selected to illustrate feedback differences observed across tasks.}
  \centering 
\includegraphics[width=0.8\textwidth,height=2\textheight,keepaspectratio]{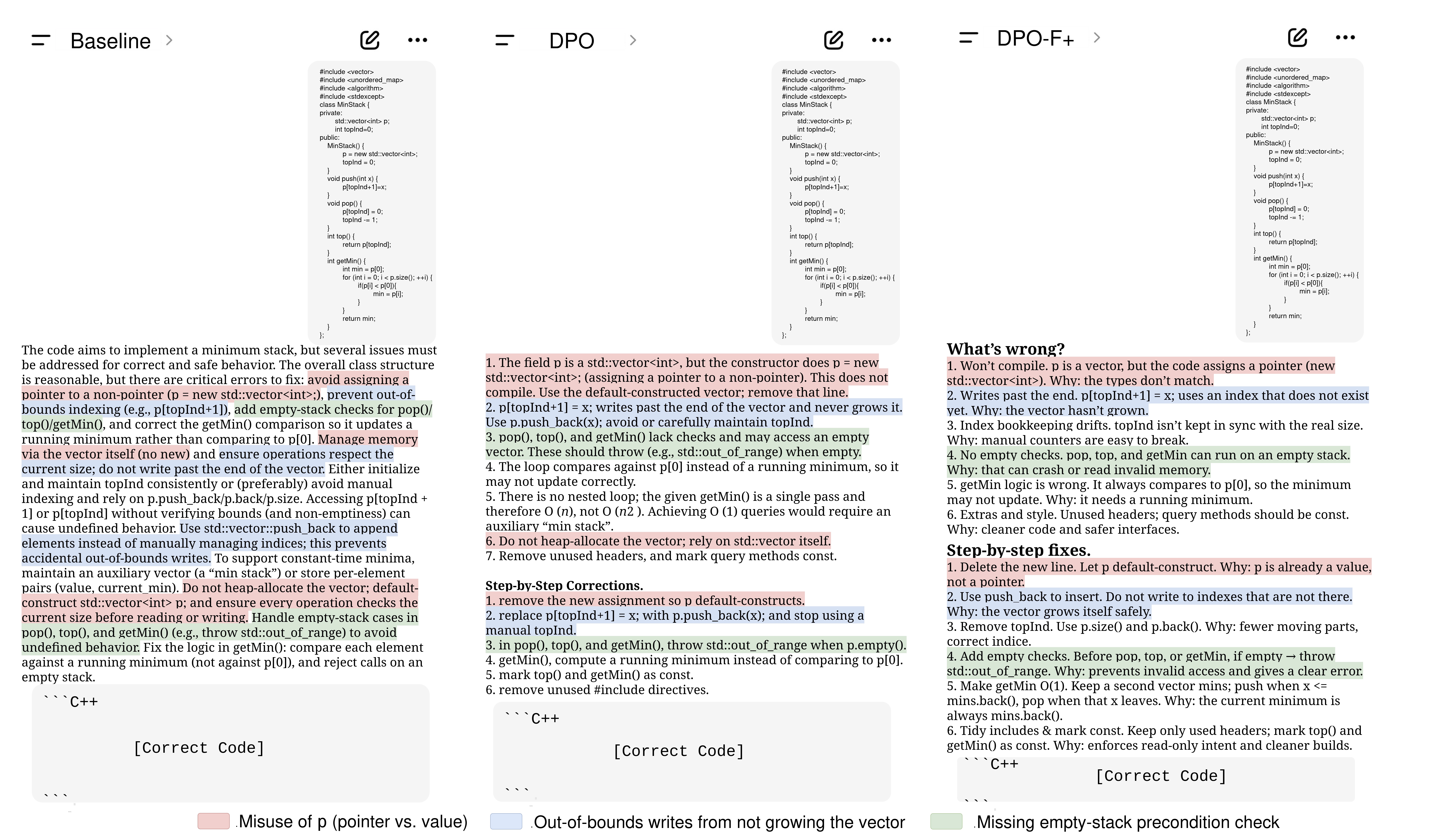}

  \label{fig:motivation}
\end{figure*}

\section{Motivation Example}
\label{sec:motivation}
In software engineering, developers increasingly rely on LLMs not only to repair code but also to explain what to fix and why. Yet the feedback accompanying LLM-generated code is often hard to follow~\cite{zi2025would,chen2025empirical}. Prior work shows that such feedback is frequently generic, under-specified, and weakly justified, making it difficult to act on, especially for novices~\cite{lohr2025you}. Without clear guidance on what to change, why the change resolves the bug, and how to apply edits in sequence, learners may introduce new faults. Prompting skill also strongly affects novice outcomes, highlighting the need for better alignment~\cite{fang2025comparative}. Similar concerns appear among professional developers, who also ask for clearer and more actionable explanations~\cite{khojah2024beyond}. To address this gap, we augment DPO with a graded preference signal, since standard DPO captures only binary pairwise orderings and ignores preference margins~\cite{rafailov2023direct}.

Figure~\ref{fig:motivation} illustrates this motivation with a buggy C++ implementation of \texttt{MinStack}. The snippet contains several design and safety flaws, including incorrect use of \texttt{p}, out-of-bounds writes from failing to grow the vector, missing precondition checks, and incorrect minimum-tracking logic. We focus on the first three issues highlighted in Figure~\ref{fig:motivation}. Although all three models identify these defects and generate correct repaired code, their feedback differs substantially in specificity and structure.

The baseline model identifies all three issues and suggests plausible fixes, but its feedback is verbose, repetitive, and weakly organized. It interleaves symptoms with fixes, rarely explains why the changes matter, and provides no concise correction plan, reducing actionability and increasing cognitive load.

DPO improves phrasing and adds concrete directives with a numbered correction list. However, its guidance is sometimes inconsistent and still incomplete: some recommendations conflict, and key rationales, such as why pointer misuse is unsafe or how failure cases should be handled, remain underexplained. Overall, DPO better states what to change, but still underspecifies why and how to apply the fixes safely.

In contrast, \textsc{DPO-f+} pairs precise edits with clear rationale. It explains the type mismatch, replaces manual writes with \texttt{push\_back} while clarifying safe container growth, and adds explicit empty-stack guards with defined failure behavior. Its guidance is stepwise, compact, and logically ordered, making each action concrete, verifiable, and easier to transfer to new contexts.

\section{Related Work}
\subsection{Reinforcement Learning from Human Feedback}
Reinforcement Learning from Human Feedback (RLHF) has emerged as a promising approach for aligning LLMs with human preferences~\cite{chaudhari2024rlhf}. 
Early work such as InstructGPT~\cite{ouyang2022training} showed that, despite remaining limitations, fine-tuning with human feedback is an effective direction for aligning language models with human intent.
Traditional RLHF typically uses Proximal Policy Optimization (PPO)~\cite{schulman2017proximal} together with a learned reward model. PPO stabilizes training by constraining policy updates, while the reward model provides the optimization signal~\cite{schulman2017proximal}. 
However, this pipeline is computationally expensive and often difficult to stabilize.
Direct Preference Optimization (DPO) addresses these limitations by reframing preference learning as a classification problem over preferred and dispreferred responses, enabling direct policy optimization without a separate reward model or reinforcement-learning loop~\cite{rafailov2023direct}. 
As a result, DPO retains the main benefits of RLHF while reducing implementation complexity~\cite{xu2024dpo}. 
It has been applied to tasks such as code generation~\cite{zhang2025focused} and feedback alignment in educational settings, including math tutoring and teaching assistant-guided feedback~\cite{scarlatos2024improving,woodrow2025improving}.
Despite these advances, the use of RLHF, particularly DPO, to generate high-quality feedback under profile-specific criteria in software engineering remains underexplored. 
Moreover, standard DPO captures only binary pairwise preferences and ignores preference strength, which may weaken the supervision signal and reduce data efficiency~\cite{wualphadpo}. 
To address these gaps, we propose a lightweight reward-augmented DPO framework for aligning LLM-generated code-repair feedback with developer preferences.

\subsection{LLM-based Frameworks in Software Engineering}
Numerous LLM-based frameworks have been developed for various SE tasks, including code generation~\cite{chen2021evaluating,fried2022incoder}, program repair~\cite{fan2023automated}, automated documentation~\cite{yang2024swe}, and code review~\cite{lu2023llama,tang2024codeagent}. 
These approaches typically involve fine-tuning pre-trained language models on domain-specific corpora or task-oriented datasets to improve performance on downstream SE applications.
However, these frameworks largely overlook human factors in the design and alignment of model outputs. 
Prior work highlights that human-centered considerations are critical in software engineering tools, as developers rely on interpretability and trust to use automated assistance effectively~\cite{amershi2019guidelines,barke2023grounded}.
Recent work in educational settings has applied DPO to align LLM-generated feedback with instructor preferences, resulting in feedback that is more accurate, more insightful, and preferred over state-of-the-art models (e.g., GPT-4o)~\cite{woodrow2025improving}; however, these efforts have not targeted software engineering tasks.
Therefore, a gap remains in applying preference-based alignment methods (e.g., DPO) to software engineering settings where feedback must be technically accurate, actionable, and tailored to developers with varying levels of experience. This work addresses that gap by proposing a feedback-alignment framework for code repair in SE contexts using profile-specific criteria.

\subsection{Code Feedback in Software Engineering}
Providing effective feedback on code is a fundamental component of software development, supporting activities such as code review~\cite{turzo2023towards}, testing~\cite{grechanik2012automatically}, software maintenance~\cite{majumdar2022automated}, and programming education~\cite{messer2024automated}. Traditional feedback systems have primarily relied on static or dynamic analysis~\cite{arifi2015automatic} and rule-based approaches~\cite{koziolek2020rule}.
With the emergence of LLMs, recent work has explored their potential for generating code feedback. For example, LLMs have been used to explain compiler errors~\cite{widjojo2023addressing}, demonstrating their utility as real-time debugging aids; to produce code review comments~\cite{tufano2022using}, often matching or surpassing the performance of heuristic and rule-based methods; and to suggest edits for improving code quality~\cite{liu2024automated} through frameworks that can effectively identify and address issues such as poor naming, code smells, and anti-patterns.
Despite these advances, most systems are instruction-tuned on general-purpose datasets or optimized for surface-level metrics (e.g., BLEU, ROUGE), rather than explicitly aligned with developer needs for feedback usefulness, correctness, and clarity. 
Given that software development is fundamentally human-centric~\cite{grundy2020towards}, such alignment is essential because developer expertise, workload, and emotional state affect how developers process feedback~\cite{razzaq2024systematic}, while tone, trust, and team dynamics shape how it is received~\cite{kortum2019behavior}.
As human–AI teaming becomes increasingly common in software engineering~\cite{mastropaolo2023robustness}, especially for coding, it is crucial to design AI-generated feedback frameworks that not only improve code correctness but also align with developer preferences across experience levels. 
Yet few alignment pipelines have been built with these requirements in mind.
To address this gap, we introduce a DPO-inspired framework that aligns LLM-generated feedback with developer needs and expectations in code-repair settings.
\begin{figure}[t]
    \centering
    \caption{The overview of \textsc{DPO-f+} framework.}    \includegraphics[width=0.8\linewidth]{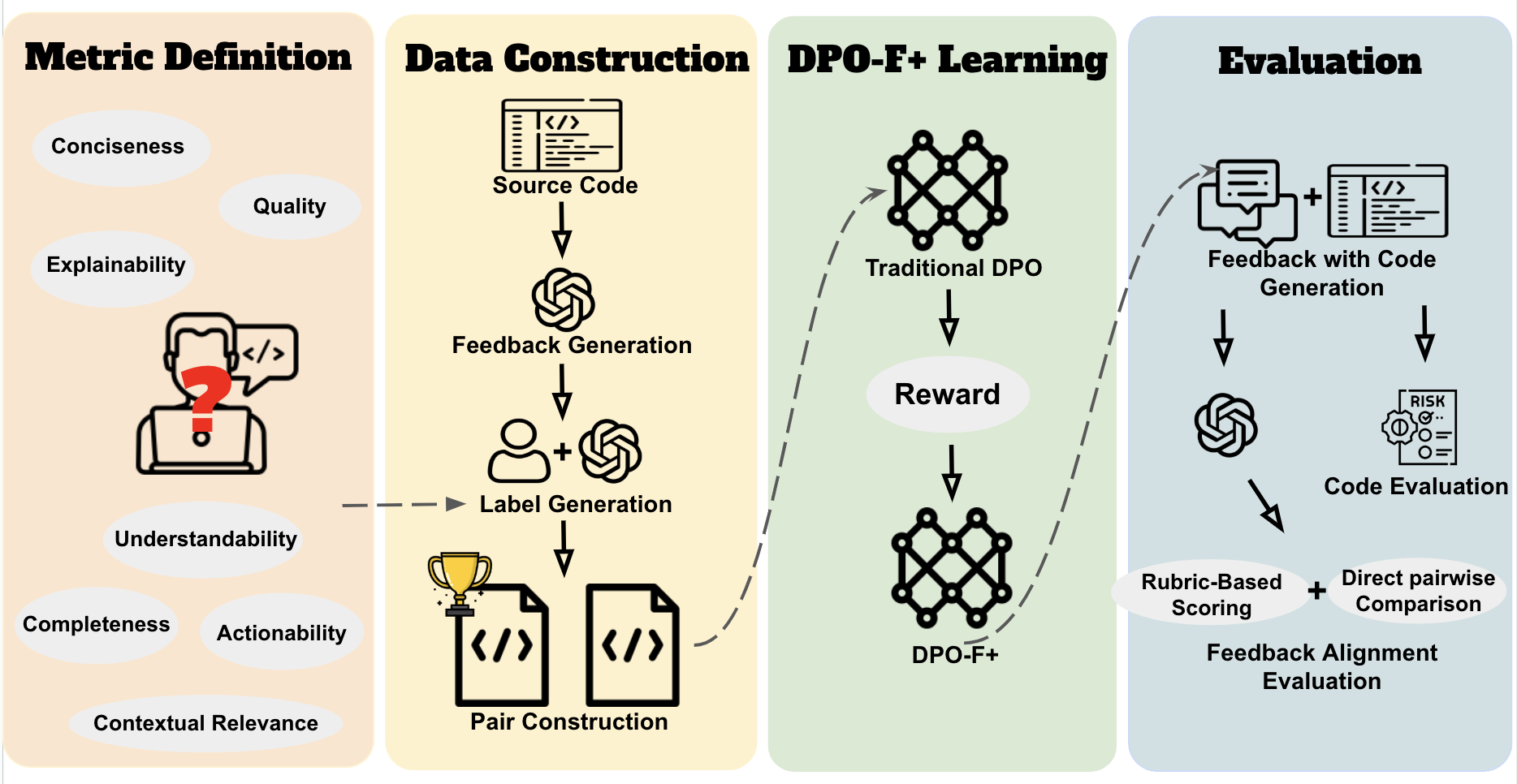}
    \label{fig:overview}
\end{figure}
\section{Methodology}
In this section, we describe the proposed \textsc{DPO-f+} framework, as summarized in Figure~\ref{fig:overview}.
Specifically, we first define the alignment metrics and use them to construct a pairwise preference dataset, then introduce a reward-augmented DPO training procedure, and finally describe the evaluation protocol.
We aim to answer the following research questions:
\begin{enumerate}
\item Does \textsc{DPO-f+} improve the alignment of code-repair feedback for novice programming instruction compared to Baseline and DPO?
\item Does \textsc{DPO-f+} maintain strong performance across more challenging and diverse code-repair tasks?
\end{enumerate}

\subsection{Metric Definition}
\label{sec:label}
To evaluate whether generated feedback matches developers’ preferences, we introduce a seven-dimensional metric for code repair guidance. Grounded in prior work, the metric operationalizes feedback qualities that developers consistently value in code reviews~\cite{bosu2015characteristics,yang2023evacrc,pascarella2018information}: \textit{Conciseness} (brief and free of unnecessary detail), \textit{Quality} (technically correct and aligned with best practices), \textit{Explainability} (clearly motivates each suggested change), \textit{Understandability} (well-structured and easy to follow), \textit{Completeness} (addresses required fixes and relevant edge cases), \textit{Actionability} (immediately usable with minimal interpretation), and \textit{Contextual Relevance} (includes enough context to clarify when and why the fix applies).

These dimensions provide a general baseline, which we adapt to reflect differences in developer experience and task demands. For novice-oriented repairs, we emphasize accessible language, step-by-step guidance, safe patterns, copy-pasteable patches with lightweight checks, and clearly stated scope. For experienced-oriented repairs, we prioritize concise prose with precise code anchors, idiomatic and composable edits, brief rationale with trade-offs, broader edge-case coverage, tool-supported verification and rollback, and consistency with the system’s architecture and deployment constraints.
An example of how we tailor the description of each dimension to different profiles of developers and task needs is summarized in Table~\ref{tab:feedback-metrics}.

\subsection{Data Preparation}
\label{sec:data-prep}
We construct the training dataset through source-code collection, augmentation, feedback generation, and pair construction.
\subsubsection{Source code and augmentation.} The dataset consists of human-written solutions to three introductory programming problems (TwoSum, MinStack, and TicTacToe), contributed by 53 novice programmers and made publicly available by prior work~\cite{ma2024breaking}.
Among these, 30 participants submitted solutions for TwoSum and MinStack, and 23 for TicTacToe. 
To expand the original code script, we apply augmentation methods grounded in prior research on code synthesis~\cite{jain2020contrastive}.
Specifically, we randomly performed code compression (e.g., reformatting, type up-conversion, dead-code elimination), and identifier modification (e.g., systematic variable renaming, identifier mangling). 
Following augmentation, the number of unique scripts (deduplicated by code content) increased from 53 to 534 across the three problems. 
Importantly, these augmentations are designed to maintain the original failure mode while varying surface structure, ensuring that augmentation does not systematically shift the distribution of error types.

\subsubsection{Feedback generation.} We collected natural-language code-repair feedback by prompting three state-of-the-art LLMs (GPT-4o, Claude 3.5 Sonnet, and DeepSeek-R1) on each code snippet. To diversify the feedback, we used three distinct prompts per snippet for each model. 
To support accuracy evaluation, we also instructed the models to produce a revised implementation that instantiated their recommendations. 
This yields, for each sample, a pool of candidate feedback-and-revision responses for downstream comparisons between higher- and lower-aligned guidance. 
We did not perform automated prompt tuning; the three prompts were manually drafted once and used unchanged across models and splits. 
We do not use external tools or function-calling in generation or judging. 
Unless otherwise noted, for candidate generation during dataset construction, decoding uses provider defaults (i.e., temperature or top-p are not overridden). For the final evaluation, we use fixed sampling parameters, which will be explained later.
\subsubsection{Pair construction.} Each feedback candidate, comprising natural-language guidance and corresponding revised code, is assessed on two dimensions.\\
\textbf{Feedback accuracy.} To evaluate feedback accuracy, we prompted the model to generate a revised program based on its suggestions. The extracted code was evaluated via a standardized C++ execution pipeline within an isolated environment. Programs were compiled using the C++17 standard with $O2$ optimization before being subjected to a suite of automated unit tests.
Submissions with compilation errors, runtime errors, or failed tests are marked as failures; submissions with missing or unparsable code are excluded. 
To prevent leakage, prompts contain only the problem statement and the submitted code; the test harness is applied post-generation and is not visible to the model. 
All executions are isolated (fresh workspace per run) with no network access and enforced time limits. 
In addition, we manually evaluated the feedback text itself for whether it correctly identifies the issue and proposes a valid fix.\\
\textbf{Feedback alignment.}
Using the metric in Section~\ref{sec:label}, we determine which generated feedback best matches a given developer profile and task. 
To reduce evaluation cost and latency and to mitigate bias, we use GPT-4 to score each feedback instance on all metrics, consistent with evidence that LLM judgments closely track expert ratings~\cite{wang2025can,zheng2023judging}. 
The prompts are provided in the supplementary file\footnote{\url{https://anonymous.4open.science/r/dpo_f-D1D0}}.
To further validate the automatic scores, we randomly selected 100 feedback cases for human review and obtained 95.0\% inter-annotator agreement, indicating high reliability.
A feedback instance is labeled accepted (i.e., preferred) if its revised code executes successfully, passes all unit tests, and achieves an average quality score of at least 4.0 across the seven metrics; otherwise, it is labeled rejected.
Using these criteria, we constructed a corpus of 6,284 accepted–rejected pairs. To prevent data leakage, we split at the level of source submissions: all pairs derived from the same source script (including all its augmented variants) were assigned to the same subset. We then partitioned these source groups into training (85\%), validation (5\%), and test (10\%) subsets.

\begin{table*}[t]
\centering
\caption{Feedback alignment evaluation metrics with persona-specific descriptions for novice and experienced developers. Each dimension is scored on a 1--5 Likert scale (1 = low, 5 = high).}
\label{tab:feedback-metrics}
\scriptsize
\begin{adjustbox}{max width=\textwidth}
\begin{tabular}{@{} p{0.14\textwidth} p{0.43\textwidth} p{0.43\textwidth} @{}}
\toprule
\textbf{Metric} & \textbf{Novice-oriented} & \textbf{Experienced-oriented} \\
\midrule
Conciseness &
Uses simple words and short sentences; avoids jargon and branches. &
Maximize signal-to-noise; minimal prose; diff-first; omit obvious context. \\
Quality &
Technically correct and prefers safe, beginner-friendly patterns; avoids clever tricks. &
Correct, robust, idiomatic, and composable within the existing codebase. \\
Explainability &
Plain-language reason for each change; one-sentence ``why this works'' per step. &
Short design rationale with key trade-offs and reason for selection. \\
Understandability &
Small, linear steps with exact file/line or code highlights. &
Precise pointers to file/symbol/line; patch-style references. \\
Completeness &
Fixes the bug plus common beginner pitfalls; includes basic validation and edge cases. &
Pre/post-conditions, edge cases, compatibility notes, and failure modes. \\
Actionability &
Copy-pasteable code and a quick verification step with expected output. &
Tooling-integrated apply/verify steps (test/lint/CI) plus rollback plan. \\
Contextual Relevance &
States language/framework/version and scope where the fix applies. &
Consistent with architecture, performance/observability constraints, and deployment. \\
\bottomrule
\end{tabular}
\end{adjustbox}
\vspace{-0.5em}
\end{table*}

\newcommand{\ph}[1]{\phbox{\texttt{#1}}}

\subsection{\textsc{DPO-f+} Learning}
\label{sec:ra-dpo}

DPO learning provides a reward-free mechanism for aligning models with human preferences~\cite{rafailov2023direct}. 
It operates by comparing the relative likelihoods of preferred ($y^+$) and rejected ($y^-$) responses under a policy $\pi_\theta$ and a frozen reference model $\pi_{\text{ref}}$. 
Although effective, standard DPO treats all preference pairs equally and ignores preference margins, diffusing learning across noisy signals and leading to slower training and less reliable feedback.
To improve, we augmented DPO with an auxiliary reward signal and a fixed-weight integration scheme.
\subsubsection{Policy Loss.}
We train the policy with a DPO-style pairwise log–sigmoid objective that contrasts policy and reference margins, with an explicit KL penalty to keep the policy close to a frozen reference:
\begin{equation} \begin{aligned} \mathcal{L}_{\text{policy}} &= \mathbb{E}_{(x,y^+,y^-)}\Big[ - \log \sigma\!\Big( \underbrace{s_\pi(x,y^+)-s_\pi(x,y^-)}_{\text{policy margin}}\\ &- \underbrace{s_{\text{ref}}(x,y^+)-s_{\text{ref}}(x,y^-)}_{\text{reference margin}} \Big)\Big] \\ &\quad + \gamma\,\mathrm{KL}\!\big(\pi_\theta(\cdot\mid x)\,\|\,\pi_{\text{ref}}(\cdot\mid x)\big), \end{aligned} \end{equation}
where $x$ is the prompt; $y^+$ and $y^-$ are the preferred and rejected feedback; 
$s_\pi(x,y)=\log \pi_\theta(y\mid x)$ and $s_{\mathrm{ref}}(x,y)=\log \pi_{\mathrm{ref}}(y\mid x)$ are sequence log-scores (i.e., sum of token log-probabilities); 
$\sigma(\cdot)$ denotes the logistic function; it maps a real-valued margin to $(0,1)$, which we interpret as the probability that $y^+$ is preferred to $y^-$; $\beta>0$ adjusts how sharply the loss responds to margin differences; and $\gamma>0$ sets the weight of the KL term that keeps the policy near the reference.
This loss increases the policy’s preference margin relative to the reference model while the KL term regularizes drift toward the reference.

\subsubsection{Reward Loss.}
To model preference margins explicitly, we train a lightweight reward model $r_{\phi}(x,y)$ with a logistic pairwise objective. The reward model maps each prompt--feedback pair $(x,y)$ to a single scalar score, is trained offline on the same preference pairs, and is frozen before policy optimization:

\begin{equation}
\begin{aligned}
\mathcal{L}_{\text{reward}}
&= \mathbb{E}_{(x,y^+,y^-)}\Big[
 - \log \sigma\!\Big(
 \underbrace{r_{\phi}(x,y^+) - r_{\phi}(x,y^-)}_{\text{reward margin}}
 \Big)\Big]
\;\; 
\end{aligned}
\end{equation}
where $x$ is the prompt, $y^+$/$y^-$ are the preferred/rejected feedback, with the logistic function $\sigma$ converts the score difference between the preferred and rejected responses into a probability that encodes the preference.
When the preferred response doesn’t clearly outperform the rejected one, the loss and its gradients are largest, which raises the preferred score and lowers the rejected one to make the gap clear.
The learned reward provides a scalar, graded estimate of preference strength, which we next use to inform policy updates so decisive wins influence training more than near-ties.

\subsubsection{\textsc{DPO-f+} Loss.}
We couple preference learning with an auxiliary reward by forming a reward-augmented score and applying a DPO-style margin objective against a frozen reference.
Let
\begin{equation}
\begin{aligned}
s_c(x,y)=s_{\pi}(x,y)+\lambda\,\tilde r_{\phi}(x,y),
\end{aligned}
\end{equation}
where $\tilde r_{\phi}$ is a scaled reward score and $\lambda\!\ge\!0$ is a fixed scalar hyperparameter controlling the contribution of the reward term.
The combined training loss is:
\begin{equation}
\begin{aligned}
\mathcal{L}_{\text{DPO-f+}}
&= \mathbb{E}_{(x,y^+,y^-)}\Big[
-\log \sigma\!\Big(
\underbrace{s_c(x,y^+)-s_c(x,y^-)}_{\text{combined margin}}\\
&-\underbrace{(s_{\text{ref}}(x,y^+)-s_{\text{ref}}(x,y^-))}_{\text{reference margin}}
\Big)\Big] \\
&\quad + \gamma\,\mathrm{KL}\!\big(\pi_\theta(\cdot\mid x)\,\|\,\pi_{\text{ref}}(\cdot\mid x)\big).
\end{aligned}
\end{equation}
where we replace the standard DPO score with a combined score \(s_c(x,y)=s_\pi(x,y)+\lambda\,\tilde r_\phi(x,y)\), so the loss rewards the model when the \emph{combined margin} for the preferred response exceeds the reference model’s margin.
The added \emph{reward margin} \(\lambda\,[\tilde r_\phi(x,y^+)-\tilde r_\phi(x,y^-)]\) provides a graded signal: clear wins trigger larger updates, while near-ties trigger smaller ones, reducing noise and focusing learning on actionable feedback.
During this stage, the reward \(\tilde r_\phi\) is \emph{frozen}, so only the policy \(\pi_\theta\) is updated; the reward simply shapes the DPO updates.
This keeps DPO’s \emph{reference anchoring}, improves sample efficiency by emphasizing high-confidence preferences, and avoids instability from jointly updating reward and policy.


\subsubsection{Experiment Setup}
We fine-tuned two base models (Qwen2.5-1.5B-Instruct, CodeLlama-7B-Instruct) on paired preference data using two NVIDIA A6000 GPUs.
Pairs were split 85/5/10 into train/validation/test.
We used AdamW (lr $5{\times}10^{-6}$, betas $(0.9,0.999)$, weight decay $0.1$), a cosine schedule with $3\%$ warmup, \texttt{bf16} precision, gradient checkpointing, and gradient clipping at $1.0$.
LoRA (\(r{=}16,\ \alpha{=}32,\\ \text{dropout}{=}0.05\)) was applied to the attention projections. 
Training used an effective batch size of 64 via gradient accumulation (micro-batch size 4 $\times$ 2 GPUs $\times$ accumulation 8 $=64$), and we used a max sequence length of 1024 tokens.
For DPO we used the default inverse temperature and an explicit forward-KL regularizer with $\gamma\in\{0,\,0.02\}$.
Models were trained for up to 3 epochs with early stopping and model selection based on the same validation criterion. For \textsc{DPO-f+}, the reward weight $\lambda$ was treated as a fixed training hyperparameter and selected on the validation set; it is not data-dependent at training or inference time.

\subsection{Framework Evaluation}
\label{sec:eval}
\subsubsection{Sampling.}
The framework's practical performance was evaluated on a newly constructed evaluation set that balanced original and augmented code. 
We compiled this set by first randomly drawing a 50\% sample from the original test scripts. 
The other half was generated using the same augmentation procedures described in Section~\ref{sec:data-prep}. 
After de-duplication and a compilability check to ensure code uniqueness and validity, the final evaluation set consisted of 200 C++ code snippets.

\subsubsection{Inference Setup.} 
We evaluated the baseline, DPO, and our aligned \textsc{DPO-f+} under identical decoding settings using a standardized C++ code-review prompt. 
Consistent with Section~\ref{sec:data-prep}, the prompt elicits natural-language feedback on the given snippet and requests a corrected version derived from that feedback. 
For each snippet, each model generated \(K{=}5\) candidates from the same prompt with sampling enabled (\texttt{temperature=0.7}, \texttt{top\_p=0.95}, \texttt{max\_new\_tokens=1024}), yielding \(1{,}000\) feedback instances per model. 
Unlike dataset construction (Section~\ref{sec:data-prep}), we explicitly override provider defaults and fix these decoding parameters for all models. 
For a fair comparison, all models were run with the same prompt and identical decoding settings.
Consequently, each output contains both the feedback and the corresponding corrected code, enabling the assessment of feedback alignment and feedback accuracy, which will be further discussed in the following.


\subsubsection{Feedback Evaluation.} 
We evaluated the generated responses across two dimensions: (1) \emph{feedback accuracy}, which measures the technical correctness of the implied code revisions; and (2) \emph{feedback alignment}, which assesses the quality of the natural language feedback using the metrics detailed in Section~\ref{sec:label}. This dual-faceted approach ensures that the feedback is both technically effective and well-aligned with developer requirements.\\
\textbf{Feedback Accuracy Evaluation.} Following the most common practices~\cite{chen2021evaluating,austin2021program}, we assessed the code produced from each feedback instance for \textit{executability} (i.e., whether it runs without error) and \textit{Pass@k}, the estimated probability that at least one of the $k$ independent samples passes the task’s test suite. 
The tests are adapted from the instructional materials of our institution:
\begin{itemize}
    \item \textbf{TwoSum.} A compact parameterized suite checks that the function returns \emph{distinct}, in-bounds indices that sum to the target and leaves the input array unchanged. Cases include duplicates, multiple valid pairs, negatives/zeros, and minimal sizes, and infeasible instances must throw \texttt{std::invalid\_argument}.
    \item \textbf{MinStack.} Mixed operation sequences (including duplicates, negatives, and zeros) verify the last-in, first-out behavior and that \texttt{getMin} maintains the running minimum across pushes and pops (including plateaus). Empty-stack operations must throw \texttt{std::out\_of\_range}.
    \item \textbf{TicTacToe.} Alternating legal moves on a $3\times3$ board validate win detection across rows, columns, and both diagonals. Non-terminal states and full-board draws must return \texttt{0}. Invalid actions (occupied cells or out-of-bounds) must throw \texttt{std::invalid\_\\argument} and leave the board unchanged.
\end{itemize}
To complement these automated metrics, we conducted a targeted human expert study to assess feedback-only correctness. 
We sampled 50 outputs per model (Baseline, DPO, and \textsc{DPO-f+}) and presented two expert annotators with the buggy code and the natural language feedback only (omitting the model-generated patch). 
Each item was evaluated on two criteria: \emph{diagnosis correctness} (i.e., identifying the root cause) and \emph{fix validity} (i.e., proposing a plausible resolution). 
We define the success rate as the proportion of instances satisfying both criteria. 
To ensure rigor, the annotators co-rated an initial calibration set of 30 samples, achieving an inter-rater agreement of 98.1\%.

\noindent\textbf{Feedback Alignment Evaluation.} We assessed alignment using two complementary LLM-as-a-judge procedures, following established paradigms for software engineering tasks~\cite{zheng2023judging}. To promote consistent judgments, we standardized the judge prompts, enforced a fixed output schema with an automated retry mechanism, and randomized item order to reduce position and prompt bias. The prompts used in the judging process are shown in Figure~\ref{fig:judge-prompts}. We also conducted a human study with 200 developers to evaluate their satisfaction with the generated feedback.\\
(1) \textit{Metric-based scoring:} We employed GPT-4 to rate each feedback item against the seven metrics in Table~\ref{tab:feedback-metrics} using a 1–5 scale. Scoring was performed at a temperature of 0.0 across three independent runs. We report the per-criterion score as the mean of these replicates, with the overall G-Eval score representing the mean across all metrics. \\
(2) \textit{Direct comparison:} As a robustness check, DeepSeek-V3 was used to produce deterministic pairwise judgments. Given the original code and two candidate feedback in randomized order, the judge selected \textit{A}, \textit{B}, or a \textit{Tie}. Results are reported as win, loss, and tie rates for the Baseline and DPO against \textsc{DPO-f+}.\\
(3) \textit{Human study:} We conducted a large-scale human study ($n=200$) to assess feedback alignment from the perspective of real developers. The study has been reviewed and approved by our local Institutional Review Board. Recruitment criteria required participants to be at least 18 years old, fluent in English, and have basic knowledge of C++. We collected preference judgments through a survey on Prolific\footnote{\url{https://www.prolific.com/}}, in which participants viewed buggy code along with three feedback responses generated by the Baseline, DPO, and \textsc{DPO-f+} models. 
Each participant evaluated 20 randomly sampled code-feedback sets and selected the response they preferred, yielding 4,000 judgments across 200 unique items, each evaluated an average of 20 times.


\newtcbox{\phbox}{on line,
  box align=base,
  enhanced,
  colback=white,
  colframe=black,
  boxrule=0.3pt,
  arc=0.8pt,
  boxsep=0.5pt,
  left=1pt,right=1pt,top=0.5pt,bottom=0.5pt
}

\begin{figure}[t]
\centering

\begin{subfigure}[t]{\linewidth}
\caption{Metric-based scoring}
\label{fig:judge-prompt-geval}
\begin{minipage}[t]{\linewidth}
\begin{promptboxcompact}{Judge Prompt 1: Metric-Based Scoring (GPT-4)}
\scriptsize
You are an expert code reviewer. Score the \emph{feedback} for the given persona and task on 7 metrics from 1--5. Be strict and consistent.

\textbf{Inputs:} Persona=\ph{persona}; Task=\ph{problem\_statement}; Code=\ph{submitted\_code}; Feedback=\ph{feedback\_text}.

\textbf{Metrics for novice persona:}
\begin{itemize}\itemsep1pt \parskip0pt \parsep0pt
  \item \textbf{Conciseness:} Simple and brief; avoids jargon.
  \item \textbf{Quality:} Correct and beginner-safe.
  \item \textbf{Explainability:} Clearly explains why each change works.
  \item \textbf{Understandability:} Small, linear, specific steps.
  \item \textbf{Completeness:} Fixes the bug and covers basic pitfalls.
  \item \textbf{Actionability:} Gives usable code and a quick check.
  \item \textbf{Contextual relevance:} Matches the stated language/framework/scope.
\end{itemize}

\textbf{Scale:} 5=excellent, 4=good, 3=mixed, 2=poor, 1=unacceptable.

\textbf{Rules:} Use only the inputs. Do not assume missing context. Do not produce chain-of-thought. Return JSON only.

\textbf{Output:}
\begin{verbatim}
{"scores":{"conciseness":1-5,"quality":1-5,"explainability":1-5,
"understandability":1-5,"completeness":1-5,"actionability":1-5,
"contextual_relevance":1-5}}
\end{verbatim}
\end{promptboxcompact}
\end{minipage}
\end{subfigure}

\vspace{0.25em}

\begin{subfigure}[t]{\linewidth}
\caption{Pairwise direct comparison}
\label{fig:judge-prompt-pairwise}
\begin{minipage}[t]{\linewidth}
\begin{promptboxcompact}{Judge Prompt 2: Pairwise Comparison (DeepSeek-V3)}
\scriptsize
You are an expert code reviewer. Given the same persona, task, and code, choose the better feedback overall: \textbf{A}, \textbf{B}, or \textbf{Tie}. Use the same 7 metrics as guidance. Be deterministic.

\textbf{Inputs:} Persona=\ph{persona}; Task=\ph{problem\_statement}; Code=\ph{submitted\_code}; Feedback A=\ph{feedback\_A}; Feedback B=\ph{feedback\_B}.

\textbf{Output:} Line 1: \textbf{A}, \textbf{B}, or \textbf{Tie}. Line 2: one short justification. No chain-of-thought. A/B order is randomized.
\end{promptboxcompact}
\end{minipage}
\end{subfigure}

\caption{LLM judge prompts for (a) metric-based scoring and (b) pairwise comparison.}
\label{fig:judge-prompts}
\end{figure}

\subsubsection{Generalizability}
To assess robustness and external validity beyond novice programming tasks, we evaluate our models on \textit{SWE-bench Lite}, a curated subset of \textit{SWE-bench} that comprises more challenging and diverse real-world GitHub issue--fix pairs.
We maintain the experimental pipeline described previously, adapting the prompts and alignment metrics to a professional software engineering context. To construct preference pairs for training, we randomly sampled 300 issues from the full \textit{SWE-bench} dataset, ensuring no overlap with \textit{SWE-bench Lite} to maintain strictly disjoint training and evaluation sets.
Evaluation results are reported using the official \textit{SWE-bench Lite} execution-based evaluator. Because the base models lack native retrieval capabilities for repository-scale codebases, we employ a fixed retrieval-and-analysis assistant (Claude 3.5 Sonnet) across all experiments, following the methodology of prior work~\cite{jimenez2023swe}. 
The assistant performs two specific roles: (i) re-ranking BM25 candidates and summarizing repository metadata, and (ii) condensing execution traces to isolate failure modes. Crucially, the assistant does not generate or edit code, nor does it participate in the evaluation of correctness. 
All natural-language feedback and patches are generated exclusively by the model under test. We enforce a fixed attempt budget, identical seeds, and uniform retrieval prompts to ensure that any observed performance gains are attributable to our fine-tuning approach rather than the retrieval process.

\section{Results}
We present an analysis of the framework's performance, examining its efficacy on both novice-level programming tasks and complex software engineering tasks for more experienced developers.
\subsection{RQ1: Does \textsc{DPO-f+} improve the alignment of code-repair feedback for novice programming instruction?}
\label{sec:training-outcomes}

\subsubsection{Preference Accuracy}
To evaluate the model alignment, we report preference accuracy: the fraction of test pairs for which the model assigns a higher likelihood to the preferred response than to the rejected one, given the same prompt, computed over response tokens only.
Results summarized in Table~\ref{tab:model-accuracy} show consistent gains from \textsc{DPO-f+} over both baseline and standard DPO. 
On novice programming tasks with \textit{Qwen2.5-1.5B-Instruct}, the baseline preference accuracy is \textbf{0.4511}. Applying standard DPO yields a modest increase to \textbf{0.4766} (\textbf{+2.55\,pp}). 
\textsc{DPO-f+} produces a larger gain, reaching \textbf{0.8184}, which is \textbf{+36.73\,pp} over the baseline and \textbf{+34.18\,pp} over DPO.
A similar pattern holds for \textit{CodeLlama-7B-Instruct}: DPO raises the baseline from \textbf{0.5892} to \textbf{0.6212} (\textbf{+3.20\,pp}), and \textsc{DPO-f+} achieves \textbf{0.8831} (\textbf{+29.39\,pp} over baseline; \textbf{+26.19\,pp} over DPO). 
While preference accuracy effectively reflects construct-level alignment with our metric, this metric can be susceptible to overfitting to specific metric cues. To complement this evaluation, we next analyzed the quality and alignment of the actual feedback generated by the models in practice.
\begin{table}[ht]
\centering
\small
\caption{Comparison of preference accuracy for Baseline, DPO, and \textsc{DPO-f+} on two task settings.}
\label{tab:model-accuracy}
\begin{tabular}{lccc}
\toprule
\textbf{Model / Task} & \textbf{Baseline} & \textbf{DPO} & \textbf{\textsc{DPO-f+}} \\
\midrule
\multicolumn{4}{l}{\textbf{Qwen2.5-1.5B-Instruct}} \\
\quad Novice Programming Task      & 0.4511 & 0.4766 & 0.8184 \\
\quad SWE-bench Lite   & 0.5200 & 0.5853 & 0.8055 \\
\midrule
\multicolumn{4}{l}{\textbf{CodeLlama-7B-Instruct}} \\
\quad Novice Programming Task   & 0.5892 & 0.6212 & 0.8831 \\
\quad SWE-bench Lite   & 0.5790 & 0.6550 & 0.8456 \\
\bottomrule
\end{tabular}
\end{table}
\subsubsection{Feedback Accuracy.}
We selected the \textit{CodeLlama-7B-Instruct}-based  \textsc{DPO-f+} to evaluate the accuracy of generated feedback, given its higher preference accuracy.
We evaluated the feedback accuracy through code executability and Pass@k (see Section~\ref{sec:eval}).
As summarized in Table~\ref{tab:combined-results}, executability rises from \textbf{27.9\%} (baseline) to \textbf{30.1\%} (DPO; \textbf{+2.2\,pp} over baseline) and to \textbf{36.9\%} with \textsc{DPO-f+} (\textbf{+9.0\,pp} over baseline; \textbf{+6.8\,pp} over DPO). 
Moreover, we also observed substantial and consistent gains across all Pass@k metrics.
Specifically, \textbf{Pass@1} increases from \textbf{0.0844} (baseline) to \textbf{0.1415} (\textsc{DPO-f+}; \textbf{+5.71\,pp}).
\textbf{Pass@3} more than doubles, from \textbf{0.1091} to \textbf{0.3250} (\textbf{+21.59\,pp}), and \textbf{Pass@5} rises from \textbf{0.2323} to \textbf{0.4151} (\textbf{+18.28\,pp}).
\textsc{DPO-f+} also outperforms standard DPO across the board, with absolute gains of \textbf{+3.30\,pp}, \textbf{+10.33\,pp}, and \textbf{+14.82\,pp} for \textbf{Pass@1}, \textbf{Pass@3}, and \textbf{Pass@5}, respectively.
The larger gains at higher $k$ suggest broader improvements across the candidate set rather than isolated wins, indicating that preference-based alignment translates into more reliable repair outcomes. 
In addition, to validate feedback-only correctness (see Section~\ref{sec:eval}), two authors manually annotated the feedback and observed success rates (i.e., the proportion of feedback that both identifies the root cause and proposes a plausible fix) of \textbf{0.50} for the baseline, \textbf{0.57} for DPO, and \textbf{0.66} for \textsc{DPO-f+}.

\begin{table}[t]
\centering
\caption{Comparison of overall executability and Pass@k results for the Baseline, DPO, and \textsc{DPO-f+}.}
\label{tab:combined-results}
\setlength{\tabcolsep}{5pt}
\renewcommand{\arraystretch}{0.98}
\begin{tabular}{lrrrr}
\toprule
Model & Executable (\%) & Pass@1 & Pass@3 & Pass@5 \\
\midrule
Baseline        & 27.9 & 0.0844 & 0.1091 & 0.2323 \\
\textsc{DPO}    & 30.1 & 0.1085 & 0.2217 & 0.2669 \\
\textsc{DPO-f+} & 36.9 & 0.1415 & 0.3250 & 0.4151 \\
\bottomrule
\end{tabular}
\end{table}

\subsubsection{Feedback Alignment.}
(1) \textit{LLM-as-a-Judge.} To quantify alignment in practice, we assessed the generated feedback using seven metrics (see Section~\ref{sec:label}).
Table~\ref{tab:feedback-metrics-result} reports per-metric scores (1--5; higher is better), the overall G-Eval, and the direct pairwise comparison. 
On the novice programming task, \textsc{DPO-f+} leads every metric (ranging from 3.16 to 4.23) and attains the highest G-Eval of 3.79 on average (Baseline 3.09; DPO 3.18), yielding an absolute improvement of +0.70 (+22.7\% relative improvement) over Baseline and +0.61 (+19.2\% relative improvement) over DPO. 
The direct pairwise comparisons further corroborate these trends: against \textsc{DPO-f+}, Baseline wins 39.62\% of pairs (loses 59.93\%, ties 0.46\%), and DPO wins 43.18\% (loses 56.45\%, ties 0.37\%), indicating that \textsc{DPO-f+} prevails in most direct comparisons.\\
(2) \textit{Human Study.}
We recruited 200 developers via Prolific: 155 identified as male, 43 as female, and 2 as non-binary. 
Participants had a mean age of 29.7 years and an average of 3.1 years of C++ experience. Although the generated feedback is primarily intended for novice programmers performing simpler tasks, we included participants with a range of experience levels, while emphasizing less experienced developers, to reduce the risk that limited novice expertise would bias the evaluation. 
Self-reported expertise levels were beginner (90), intermediate (64), advanced (29), and expert (17). Preference for \textsc{DPO-f+} was observed across all expertise levels. In particular, expert developers preferred \textsc{DPO-f+} in 73.0\% of comparisons, followed by beginners (71.5\%), advanced participants (64.3\%), and intermediate participants (61.6\%). 
Overall, 120 of 200 developers selected \textsc{DPO-f+} as their preferred model, which was significantly above chance (exact binomial test against 50\%: $p=0.0057$). Across all three-way comparison sets, developers selected \textsc{DPO-f+} most often (54.3\%), compared with DPO (23.1\%) and the baseline (22.6\%); this difference was significant ($\chi^2(2)=791.43$, $p<0.001$).


\begin{table}[t]
\centering
\caption{Evaluation of feedback alignment. Results show scores (1--5) across seven metrics for novice programming tasks and \textsc{SWE-Bench-Lite}. Final columns report pairwise comparison (Win/Loss/Tie, \%) of Baseline and DPO against \textsc{DPO-f+}.}
\label{tab:feedback-metrics-result}

\scriptsize
\setlength{\tabcolsep}{2pt}
\renewcommand{\arraystretch}{0.92}

\resizebox{\columnwidth}{!}{%
\begin{tabular}{@{}l
S[table-format=1.2] S[table-format=1.2] S[table-format=1.2]
S[table-format=1.2] S[table-format=1.2] S[table-format=1.2]
S[table-format=1.2] S[table-format=1.2]
S[table-format=2.2] S[table-format=2.2] S[table-format=2.2]@{}}
\toprule
& \multicolumn{7}{c}{\shortstack{Feedback Alignment}}
& {G-Eval}
& \multicolumn{3}{c}{\shortstack{vs.\ \textsc{DPO-f+}(\%)}} \\
\cmidrule(lr){2-8}\cmidrule(lr){9-9}\cmidrule(lr){10-12}
Model
& {Conc.} & {Qual.} & {Expl.} & {Und.} & {Compl.} & {Act.} & {Ctxt.}
& {Avg}
& {Win} & {Loss} & {Tie} \\
\midrule
\multicolumn{12}{c}{\textit{Novice Programming Task}}\\
\midrule
Baseline        & 3.09 & 3.30 & 2.61 & 3.61 & 2.69 & 3.14 & 3.16 & 3.09 & 39.62 & 59.93 & 0.46 \\
DPO             & 3.16 & 3.21 & 2.82 & 3.70 & 2.77 & 3.33 & 3.26 & 3.18 & 43.18 & 56.45 & 0.37 \\
\textsc{DPO-f+} & 3.95 & 4.01 & 3.16 & 4.23 & 3.34 & 4.07 & 3.79 & 3.79 & \multicolumn{1}{c}{--} & \multicolumn{1}{c}{--} & \multicolumn{1}{c}{--} \\
\midrule
\multicolumn{12}{c}{\textit{SWE-Bench-Lite}}\\
\midrule
Baseline        & 3.78 & 3.77 & 3.57 & 3.89 & 3.38 & 3.86 & 4.15 & 3.76 & 13.64 & 56.57 & 29.79 \\
DPO             & 3.74 & 3.76 & 3.67 & 3.87 & 3.39 & 3.94 & 4.20 & 3.80 & 35.92 & 52.39 & 11.69 \\
\textsc{DPO-f+} & 3.81 & 3.88 & 3.67 & 3.95 & 3.42 & 3.98 & 4.25 & 3.85 & \multicolumn{1}{c}{--} & \multicolumn{1}{c}{--} & \multicolumn{1}{c}{--} \\
\bottomrule
\end{tabular}%
}
\end{table}

\begin{table}[t]
\centering
\small
\caption{Evaluation of resolved issues on SWE-bench Lite. Patches were generated by the models and assessed using the official evaluator.}
\begin{tabular}{lrrrr}
\toprule
Model & \#Total & \#Completed & \#Resolved \\
\midrule
Baseline      & 300  & 22 &  6 \\
\textsc{DPO}   & 300  & 33 & 15 \\
\textsc{DPO-f+}& 300 & 39 &  20 \\
\bottomrule
\end{tabular}

\label{tab:swebench-lite}
\end{table}

\subsection{RQ2: Does \textsc{DPO-f+} maintain strong performance across more challenging and diverse code-repair tasks?}
\subsubsection{Preference Accuracy.}
Having established strong performance on novice tasks, we next assessed generalization to more challenging settings aimed at experienced developers. 
We reused the same data construction, training, and evaluation pipeline, making minor adjustments to the seven metric definitions and prompts to better reflect expert needs (see Section~\ref{sec:label}).

Table~\ref{tab:model-accuracy} reports preference accuracy for \textit{Qwen2.5-1.5B-Instruct} and \textit{CodeLlama-7B-Instruct} under DPO and \textsc{DPO-f+} training. 
On \textit{SWE-bench Lite}, \textit{Qwen2.5-1.5B-Instruct} rises from \textbf{0.5200} (baseline) to \textbf{0.5853} with DPO (\textbf{+6.53 pp}) and to \textbf{0.8055} with \textsc{DPO-f+} (\textbf{+28.55 pp} over baseline; \textbf{+22.02 pp} over DPO).
For \textit{CodeLlama-7B-Instruct}, DPO improves the baseline's \textbf{0.5790} to \textbf{0.6550} (\textbf{+7.60 pp}), while \textsc{DPO-f+} attains \textbf{0.8456} (\textbf{+26.66 pp} over baseline; \textbf{+19.06 pp} over DPO).
Overall, \textsc{DPO-f+} outperforms both the baseline and standard DPO, with the largest absolute gains on more challenging tasks for experienced developers.

\subsubsection{Feedback Accuracy.}
We next assessed alignment performance in a downstream setting using \textit{SWE-bench Lite}. Following the official protocol, we measured feedback accuracy via the effectiveness of the patches (i.e., whether they can resolve the corresponding issue) produced by each model. An instance is counted as \emph{Resolved} only if (i) the patch applies cleanly, (ii) makes all Fail-to-Pass tests pass, and (iii) has no Pass-to-Pass tests regress. 
Consistent with prior work, we report the number of \emph{Resolved} instances over the full evaluation split as the primary outcome, and we additionally report \emph{Completed} for transparency, where \emph{Completed} denotes the number of instances that successfully finished the evaluation procedure (irrespective of being Resolved). For a fair comparison, we used the same prompt across models to produce patch feedback and evaluated generated patches on the same 300 instances under the official Docker-based harness, restricting evaluation to one prediction (i.e., feedback) per instance.

Under this protocol, our \textsc{DPO-f+} model resolves \textbf{6.67\%} of tasks (20/300), compared with \textbf{5.00\%} (15/300) for standard DPO and \textbf{2.00\%} (6/300) for the baseline, yielding absolute gains of \textbf{+1.67 pp} over DPO and \textbf{+4.67 pp} over the baseline.
While the absolute rate is constrained by the 7B base model’s capacity, \textsc{DPO-f+} nonetheless delivers a clear, effective improvement in patch quality over standard DPO under identical retrieval support.

\subsubsection{Feedback Alignment.}
We conducted a further evaluation of alignment for model-generated natural-language feedback under more challenging task conditions.
Consistent with the novice-task results, \textsc{DPO-f+} attains the top score on all seven metrics as reported in Table~\ref{tab:feedback-metrics-result} and the highest G-Eval mean (3.85), a relative improvement of 2.39\% over the baseline (3.76) and 1.32\% over DPO (3.80).
Direct pairwise comparisons corroborate these trends: against \textsc{DPO-f+}, the baseline only wins 13.64\% of pairs (loses 56.57\%, ties 29.79\%), and DPO wins 35.92\% (loses 52.39\%, ties 11.69\%), indicating that \textsc{DPO-f+} prevails in most direct comparisons on this benchmark.
We do not conduct human evaluation 
on this benchmark, as each task requires deep repository-level context and 
domain expertise that would make it infeasible for crowd-sourced participants 
to reliably assess feedback quality.






\section{Limitations}

\subsection{Data Construction and Optimization}
Our novice-task setting used a small but authentic dataset: 53 novice submissions across three introductory C++ tasks, yielding 83 original code snippets containing bugs written by novice developers. 
We retained real novice-written code to preserve realistic error patterns, but this dataset does not cover the full range of novice mistakes. We therefore applied code augmentation~\cite{jain2020contrastive} to expand the source pool, while noting that these transformations mostly preserve existing bug semantics rather than introducing substantially new failure modes. 
To test transfer beyond this narrow instructional setting, we additionally evaluated the framework on more complex tasks.

For efficiency, we used LoRA rather than full-parameter fine-tuning and only explored 1.5B- and 7B-scale models. As a result, our novice-task and \textit{SWE-bench Lite} results are not directly comparable to state-of-the-art systems built with larger datasets, stronger backbones, or more expensive training. Our findings should therefore be interpreted as evidence that the framework improves feedback alignment in a constrained educational setting, not as a claim of broad generalization.

Finally, we compared \textsc{DPO-f+} mainly against a baseline and standard DPO, rather than stronger RL-based alternatives. We therefore do not claim optimality in exploration, sample efficiency, or robustness to distribution shift.
\subsection{Alignment Evaluation}
Beyond data and training constraints, our evaluation combines LLM-as-a-judge with a human study. We use LLM-based judging because it is practical, scalable, reproducible, and applicable to both novice tasks and the more complex \textsc{SWE-Bench-Lite} setting. Still, we treat it as a proxy rather than a substitute for real developer evaluation, since it may not fully capture how developers perceive, trust, or act on feedback in practice, nor does it directly measure effects on code comprehension. Prior work suggests that LLM-as-a-judge can achieve human-level performance on software engineering tasks under similar protocols \cite{wang2025can}. To reduce model-specific bias, we use two complementary protocols (metric-based and pairwise), a cross-family judge ensemble, anonymized and randomized pairwise presentation, deterministic decoding, score aggregation across judges, and manual auditing of a subset of judgments, yielding about 95\% inter-annotator agreement on 100 items.
To address this limitation, we also conducted a human study on novice tasks, in which feedback could be evaluated realistically by recruited participants. We did not extend the human study to \textsc{SWE-Bench-Lite}, as those tasks are substantially more complex and would require a much larger pool of qualified participants to assess reliably at scale.

\subsection{Generalizability}
The evaluation of \textsc{DPO-f+} covers novice programming tasks and the more challenging \textit{SWE-Bench-Lite}, approximating novice-oriented and experienced-oriented settings in C++ and Python. We treat the novice–experienced split as a coarse first step toward richer developer profiling (e.g., role, domain familiarity, and accessibility needs), which we leave to future work.
This scope, while representative, excludes several prominent ecosystems (e.g., Java and JavaScript) and domain-specific languages (e.g., SQL and shell). Future work could broaden language coverage by incorporating diverse open-source corpora and establishing cross-language benchmarks to assess the framework’s generalizability across programming paradigms.

Additionally, our assessment focuses on code repair and does not address API migration, test generation/repair, documentation updates, etc. 
Future work could extend the framework to these activities by developing task-specific datasets and metrics, enabling a more comprehensive evaluation of applicability across the software development lifecycle.

\section{Discussion}
\subsection{Personalized LLMs for Code Comprehension Support}
\label{sec:dis-1}
We proposed a practical framework to better align LLM-generated feedback with developers’ preferences and needs. For evaluation, we moved beyond objective accuracy and introduced customizable yet scalable quality dimensions (e.g., Conciseness, Technical Quality, Explainability, Understandability, Completeness, Actionability, and Contextual Relevance), providing a comprehensive view of natural-language feedback for code-repair tasks to help developers understand the generated content better. Framing evaluation with structured metrics (rather than open-ended preference) yields consistent, comparable scores across prompts, languages, and tasks while maintaining predictable evaluation cost. When complemented with light human validation on sampled items, we observed strong agreement between automated and human judgments. 

The framework supports profile-specific objective shaping by weighting metric dimensions to match context (e.g., assigning greater weight to Explainability for novices and to Completeness in safety-critical settings). Emphasizing specific dimensions for specific goals encourages the model to produce clearer rationales, guidance, and contextual cues that align with developers’ needs, helping them reduce the cognitive load when dealing with the generated content and localize faults more quickly. Thus, this targeted shaping of feedback may yield measurable gains on code comprehension tasks.
Future work could conduct human studies to further validate this.

Finally, since the metrics and data construction can be adjusted quickly and easily, the framework is well-suited to limited resource settings (e.g., CS education), enabling efficient feedback generation, rapid what-if evaluations of prompts/models, and scalable monitoring when human graders or labeled data are limited. Nevertheless, since preferences vary across teams and expertise levels, we still recommend a human-in-the-loop process with periodic audits to incorporate real-world signals and refine the alignment procedure.

\subsection{Failure Analysis}
\label{sec:failure-analysis}
While \textsc{DPO-f+} improves feedback alignment and repair outcomes on average (Section~\ref{sec:training-outcomes}), we also observe systematic cases where it underperforms relative to the baseline or standard DPO. To make these limitations transparent, we summarize the dominant failure modes in \emph{regression cases}, defined as instances where \textsc{DPO-f+} performs worse than the comparison model under the evaluation protocol in Section~\ref{sec:training-outcomes}.

\noindent (1) \textit{Over-verbosity and Signal Dilution.} In about 19\% of regression cases, \textsc{DPO-f+} produces substantially longer feedback than the baseline, often adding elaborations that do not improve actionability. This can increase cognitive load and obscure the main corrective action, especially for localized bugs where a concise diagnosis and targeted fix are sufficient. We conjecture that preference-based optimization may sometimes reward fluency and completeness at the expense of information density.

\noindent (2) \textit{Lack of Explicit Diagnostic Framing.} In 98\% of regression cases, the model does not explicitly frame the problem in diagnostic terms (e.g., using words such as ``bug,'' ``issue,'' ``root cause,'' or ``this fails because''). As a result, even plausible edits may underemphasize the linguistic cues that help developers recognize \emph{what is wrong} before acting.

\noindent (3) \textit{Fix-Centricity without Causal Explanation.} Relatedly, the model often prescribes edits without explaining the failure mechanism. That is, it states \emph{what to change} but not \emph{why the original code fails} or why the proposed fix works. This weakens the pedagogical value of the feedback and may reduce transfer to future problems. One mitigation is to explicitly separate \textit{Diagnosis} and \textit{Fix} during training and evaluation, for example by requiring a brief causal diagnosis followed by a minimal grounded fix.

\noindent (4) \textit{Absence of Code-Specific Anchors.} In 9\% of regression cases, the feedback lacks code-specific identifiers, such as variable names, function signatures, or references to relevant lines or blocks. Without such anchors, suggestions are harder to map to the code and easier to misapply. Strengthening anchoring requirements in the alignment metric is a promising direction for improving grounding.

Therefore, DPO-f+ is preferred overall, but its remaining weakness is that it can still produce feedback that is fluent and helpful-sounding without being diagnostic, grounded, and concise enough for real code understanding.

\subsection{Future Applications and Integration}
\label{sec:dis-2}
Our failure analysis suggests that effective deployment of \textsc{DPO-f+} requires application-specific constraints to reduce regression modes. In particular, four needs emerge: (i) \textit{verbosity control} to prevent signal dilution; (ii) \textit{explicit diagnostic framing} to state the cause before prescribing edits; (iii) \textit{structured separation} between diagnosis and fix to preserve pedagogical value; and (iv) \textit{code-specific anchoring} to ground suggestions in the source. These findings motivate future work that combines refined alignment metrics with interface affordances, such as sectioned outputs and length budgets, to improve robustness across settings.

\noindent (1) \textit{Open-Source Development.} \textsc{DPO-f+} could be adapted as a review assistant that improves review throughput without increasing triage burden. By mining preference pairs from repository traces, such as pull-request discussions and CI outcomes, the framework could be optimized for \emph{Actionability} and \emph{Contextual Relevance}. To reduce the failure modes identified in Section~\ref{sec:failure-analysis}, such an assistant should enforce lightweight constraints, including identifier mentions and bounded-length fix plans, so suggestions remain concrete and transferable across unfamiliar codebases.

\noindent (2) \textit{Computer Science Education.} In educational settings, \textsc{DPO-f+} could serve as a scalable TA-style assistant. Unlike standard LLMs that may encourage ``answer-dumping,'' the \textsc{DPO-f+} objective can be weighted toward \emph{Understandability} and \emph{Explainability}. A diagnosis-first structure that states \textit{what} is wrong and \textit{why} it fails may better target student misconceptions and reduce the cognitive load of interpreting raw code patches. Evaluation in this setting should extend beyond execution correctness to include learning outcomes, such as time-to-first-pass and performance on follow-up tasks.

\noindent (3) \textit{Collaborative Software Teams.} For professional teams, \textsc{DPO-f+} could be aligned with internal architectural standards and historical review practices. To reduce misapplication in complex systems, future versions should prioritize \emph{Completeness} and \emph{Contextual Relevance}, requiring references to affected modules or APIs and concrete code anchors when possible. Incorporating such grounding requirements into the alignment metric may help organizations use AI feedback to reduce defects while maintaining coding standards.

Across these domains, moving from surface-level fluency to structured, diagnostic feedback is an important step toward making AI-driven code review more reliable and pedagogically sound.

\section{Conclusion}
Large language models are increasingly used for software engineering tasks such as code repair. However, developers still struggle to interpret model outputs, which limits effective human–AI teaming. Prior work has largely focused on optimizing generated code while giving less attention to the natural-language feedback that can support code comprehension and iterative improvement. To address this gap, we present \textsc{DPO-f+}, a framework that aligns code-repair feedback with developer needs through profile-specific criteria to better support code comprehension. The framework (1) formalizes profile- and domain-specific metrics for feedback alignment, (2) automatically constructs preference pairs from code-repair tasks, (3) fine-tunes models using Direct Preference Optimization (DPO) with a reward, and (4) evaluates performance at scale through an automated protocol.
Empirically, \textsc{DPO-f+} outperforms both the baseline and standard DPO in generated-code accuracy and feedback alignment. On novice programming tasks, it improves Pass@1 by 5.71 percentage points over the baseline and 3.30 points over standard DPO. On more advanced tasks (\textit{SWE-bench Lite}), it improves the issue-resolution rate by 4.67 points over the baseline and 1.67 points over DPO. Across both settings, it also achieves the highest overall feedback-alignment scores. We further conducted a human study ($n=200$), in which developers reported greater satisfaction with feedback generated by \textsc{DPO-f+}. By aligning feedback to developer needs, \textsc{DPO-f+} reframes LLM-assisted repair from one-shot output delivery into a collaborative sense-making process, offering a practical path toward improved code comprehension and human–AI teaming in software engineering.

\section{Data Availability}
Our source data and scripts are available at \url{https://doi.org/10.5281/zenodo.19337617}.
\clearpage
\bibliographystyle{ACM-Reference-Format}
\bibliography{my}

\end{document}